\documentclass[a4paper,12pt]{article}
\usepackage[english]{babel}
\usepackage{jheppub2}
\usepackage{subfig}
\usepackage[T1]{fontenc} 
\usepackage{graphicx}
\usepackage{epsfig}
\usepackage{amsmath}
\usepackage{amssymb}
\usepackage{float}
\usepackage{placeins}
\usepackage{braket}
\usepackage{slashed}
\usepackage{mathdots}
\usepackage{lipsum}

\allowdisplaybreaks[4]

\author[a]{Marco Bochicchio}
\affiliation[a]{Physics Department, INFN Roma1,\\
Piazzale A. Moro 2, Roma, I-00185, Italy}

\emailAdd{marco.bochicchio@roma1.infn.it}

\abstract
{We revisit the operator mixing in massless QCD-like theories. In particular, we address the problem of determining under which conditions a renormalization scheme exists where the renormalized mixing matrix in the coordinate representation, $Z(x, \mu)$, is diagonalizable to all perturbative orders. As a key step, we provide a differential-geometric interpretation of renormalization that allows us to apply the Poincar\'e-Dulac theorem to the problem above: We interpret a change of renormalization scheme as a (formal) holomorphic gauge transformation, $-\frac{\gamma(g)}{\beta(g)}$ as a (formal) meromorphic connection with a Fuchsian singularity at $g=0$, and $Z(x,\mu)$ as a Wilson line, with $\gamma(g)=\gamma_0 g^2 + \cdots$ the matrix of the anomalous dimensions and $\beta(g)=-\beta_0 g^3 +\cdots$ the beta function. As a consequence of the Poincar\'e-Dulac theorem, if the eigenvalues $\lambda_1, \lambda_2, \cdots $ of the matrix $\frac{\gamma_0}{\beta_0}$, in nonincreasing order $\lambda_1 \geq \lambda_2 \geq \cdots$, satisfy the nonresonant condition $\lambda_i -\lambda_j -2k  \neq 0$ for $i\leq j$ and $k$ a positive integer, then a renormalization scheme exists where $-\frac{\gamma(g)}{\beta(g)} = \frac{\gamma_0}{\beta_0} \frac{1}{g}$ is one-loop exact to all perturbative orders. If in addition $\frac{\gamma_0}{\beta_0}$ is diagonalizable, $Z(x, \mu)$ is diagonalizable as well, and the mixing reduces essentially to the multiplicatively renormalizable case. We also classify the remaining cases of operator mixing by the Poincar\'e-Dulac theorem.}

\newcommand{\beq}{\begin{equation}}

\newcommand{\eeq}{\end{equation}}

\newcommand{\bea}{\begin{eqnarray}}
\newcommand{\eea}{\end{eqnarray}}
\newcommand{\bfig}{\begin{figure}}
\newcommand{\efig}{\end{figure}}
\newcommand{\bc}{\begin{center}}
\newcommand{\ec}{\end{center}}

\title{On the geometry of operator mixing in massless QCD-like theories}
\date{}
\begin{document}
\maketitle
\flushbottom

\section{Introduction and physics motivations}

In the present paper we revisit the operator mixing in asymptotically free gauge theories massless to all perturbative orders, such as QCD with massless quarks. We refer for short to such theories as massless QCD-like theories.\par
In fact, nonperturbatively, according to the renormalization group (RG), massless QCD-like theories develop a nontrivial dimensionful scale that labels the RG trajectory -- the RG invariant -- $\Lambda_{RGI}$:
\bea \label{1}
 \Lambda_{RGI}  \sim   \mu \, e^{-\frac{1}{2\beta_0 g^2}} g^{-\frac{\beta_1}{ \beta_0^2}} c_0 (1+\sum^{\infty}_{n=1} c_n g^{2n})
 \eea
-- the only free parameter \cite{MBR,MBL} in the nonperturbative S matrix of confining massless QCD-like theories \cite{MBR,MBL} -- that any physical mass scale must be proportional to, with $\beta_0$ and $\beta_1$ the renormalization-scheme independent first-two coefficients of the beta function $\beta(g)$:
\bea \label{2}
\frac{\partial g}{\partial \log \mu}=\beta(g)= -\beta_0 g^3 - \beta_1 g^5 + \cdots 
\eea
and $g=g(\mu)$ the renormalized coupling.\par
Hence, our main motivation is for the study of the ultraviolet (UV) asymptotics, implied by the RG, of $2$-, $3$- and $n$-point correlators of gauge-invariant operators for the general case of operator mixing, in relation to an eventual nonperturbative solution, specifically in the large-$N$ limit \cite{H,V,Migdal,W}.\par
In this respect, the study of the UV asymptotics for correlators of multiplicatively renormalizable operators \cite{MBM,MBN,MBH}, apart from the intrinsic interest \cite{R}, sets powerful constraints \cite{MBM,MBN,MBH,MBR,MBL,BB} on the nonperturbative solution of large-$N$ confining QCD-like theories.\par
Accordingly, the present paper is the first of a series, where we intend to study the structure of the UV asymptotics of gauge-invariant correlators implied by the RG in the most general case above, in order to extend the aforementioned nonperturbative results \cite{MBM,MBN,MBH,MBR,MBL,BB} to operator mixing.\par
In particular, since operator mixing is ubiquitous in gauge theories, an important problem, which is hardly discussed in the literature, is to determine under which conditions it may be reduced, to all orders of perturbation theory, to the multiplicatively renormalizable case. \par
The aim of the present paper is to solve this problem, and also to classify the cases of operator mixing where the aforementioned reduction is not actually possible. \par

\section{Main results and plan of the paper}

We can exemplify the structure of the UV asymptotics of $2$-point correlators as follows. 
 In massless QCD-like theories, we consider $2$-point correlators in Euclidean space-time: 
\bea
G_{ik}(x) = \langle O_i(x) O_k(0) \rangle 
\eea
of renormalized local gauge-invariant operators $O_i(x)$:
\bea
O_i(x)= Z_{ik} O_{Bk}(x)
\eea
where $O_{Bk}(x)$ are the bare operators that mix \footnote{In fact \cite{M01,M02,M03}, gauge-invariant operators also mix with BRST-exact operators and with operators that vanish by the equations of motion (EQM). But correlators of gauge-invariant operators with BRST-exact operators vanish, while correlators with EQM operators reduce to contact terms. Hence, for our purposes it suffices to take into account the mixing of gauge-invariant operators only.} under renormalization and $Z$ is the bare mixing matrix.\par
The corresponding Callan-Symanzik equation \cite{C,S,Pes,Zub} reads in matrix notation \cite{BB2}:
\begin{equation}
\label{2.1}
\left(x \cdot \frac{\partial}{\partial x}+\beta(g)\frac{\partial}{\partial g}+2D\right)G+\gamma(g) \, G+G \, \gamma^T(g)=0
\end{equation}
with $\gamma^T$ the transpose of $\gamma$, $D$ the canonical dimension of the operators, and $\gamma(g)$ the matrix of the anomalous dimensions \footnote{The sign of the coefficient matrices in eq. \eqref{1.6}, $\gamma_0, \gamma_1, \cdots$, is the standard one, but opposite with respect to the convention employed in \cite{MBM,MBN,MBH,MBR,MBL,BB}.}:
\begin{equation}
\label{1.6}
\gamma(g)=- \frac{\partial Z}{\partial \log \mu} Z^{-1}=  \gamma_{0} g^2  + \gamma_{1} g^4 +\cdots
\end{equation}
The general solution has the form:
\begin{equation}
\label{2.2}
G(x) =
Z(x, \mu)\mathcal{G}(x,g(\mu),\mu)Z^T(x, \mu)
\end{equation}
with $\mathcal{G}(x,g(\mu),\mu)$ satisfying: 
\begin{equation}
\label{2.3}
\left(x \cdot \dfrac{\partial}{\partial x}+\beta(g)\dfrac{\partial}{\partial g}+2D \right)\mathcal{G} = 0
\end{equation}
and:
\begin{equation}
\label{01.90}
Z(x, \mu)=P\exp\left(-\int^{g(\mu)}_{g(x)}\frac{\gamma(g)}{\beta(g)}dg\right)
\end{equation}
where $Z(x, \mu)$ is the renormalized mixing matrix in the coordinate representation, $P$ denotes the path ordering of the exponential, and $g(\mu)$, $g(x)$ are short notations for the running couplings $g(\frac{\mu}{\Lambda_{RGI}})$, $g(x \Lambda_{RGI})$ at the corresponding scales, with UV asymptotics:
\begin{equation}
\label{1.12}
g^2(x \Lambda_{RGI})  \sim \dfrac{1}{\beta_0\log(\frac{1}{x^2 \Lambda_{RGI}^2})} \left(1-\dfrac{\beta_1}{\beta_0^2} \dfrac{\log\log(\frac{1}{x^2 \Lambda_{RGI}^2})}{\log(\frac{1}{x^2 \Lambda_{RGI}^2})}\right)
\end{equation}
We will discuss the UV asymptotics of $\mathcal{G}(x,g(\mu),\mu)$ in \cite{BB2}, while
in the present paper we concentrate on the UV asymptotics \footnote{In the present paper $\gamma(g)$ and $\beta(g)$ in eq. \eqref{01.90} are actually only defined in perturbation theory by eqs. \eqref{2} and \eqref{1.6}. In this case, eq. \eqref{01.90} only furnishes the UV asymptotics of $Z(x, \mu)$, thanks to the asymptotic freedom.} of $Z(x, \mu)$.\par
In the general case, because of the path-ordered exponential and the matrix nature of eq. \eqref{01.90}, it is difficult to work out the actual UV asymptotics of $Z(x, \mu)$. \par
Of course, were $\frac{\gamma(g)}{\beta(g)}$ diagonal, we would get immediately the corresponding UV asymptotics for $Z(x, \mu)$, as in the multiplicatively renormalizable case \cite{Pes,Zub}.\par
Therefore, the main aim of the present paper is to find under which conditions a renormalization scheme exists where $Z(x, \mu)$ is diagonalizable to all perturbative orders.\par
Another aim is to classify the cases of operator mixing where such a diagonalization is not possible.\par
We accomplish the aforementioned purposes in three steps: \par
In the first step (section \ref{3}), we furnish an essential differential-geometric interpretation of renormalization: We interpret a change of renormalization scheme as a (formal) holomorphic gauge transformation, $-\frac{\gamma(g)}{\beta(g)}$ as a (formal) meromorphic connection with a Fuchsian singularity at $g=0$, and $Z(x,\mu)$ as a Wilson line. \par
In the second step (section \ref{4}), we employ the above interpretation 
to apply in the framework of operator mixing -- for the first time, to the best of our knowledge -- the theory of canonical forms, obtained by gauge transformations, for linear systems of differential equations with meromorphic singularities \cite{PD0}, and specifically (a formal version of) the Poincar\'e-Dulac theorem \cite{PD1} for Fuchsian singularities, i.e., simple poles. \par
In the third step (section \ref{PD}), we provide a condensed proof of the Poincar\'e-Dulac theorem in the case (I) below, where $Z(x,\mu)$ is diagonalizable to all orders of perturbation theory.\par
From the three steps above, our conclusions follow:\par
As a consequence of the Poincar\'e-Dulac theorem, if the eigenvalues $\lambda_1, \lambda_2, \cdots $ of the matrix $\frac{\gamma_0}{\beta_0}$, in nonincreasing order $\lambda_1 \geq \lambda_2 \geq \cdots$, do not differ by a positive even integer (section \ref{4}), i.e.:
\bea \label{rce}
 \lambda_i -\lambda_j -2k  \neq 0
 \eea
 for $i\leq j$ and $k$ a positive integer, then it exists a renormalization scheme where: 
 \bea \label{0}
 -\frac{\gamma(g)}{\beta(g)}= \frac{\gamma_0}{\beta_0} \frac{1}{g}
 \eea
is one-loop exact to all orders of perturbation theory, with $-\frac{\gamma(g)}{\beta(g)}$ defined in eq. \eqref{A}.
\par
Moreover, according to the terminology of the Poincar\'e-Dulac theorem, our classification of operator mixing is as follows: \par
If a renormalization scheme exists where $ -\frac{\gamma(g)}{\beta(g)}$ can be set in the canonical form of eq. \eqref{0}, we refer to the mixing as nonresonant, that by eq. \eqref{rce} is the generic case.
Otherwise, we refer to the mixing as resonant.\par
Besides, $\frac{\gamma_0}{\beta_0}$ may be either diagonalizable \footnote{A sufficient condition for a matrix to be diagonalizable is that all its eigenvalues are different.} or nondiagonalizable.\par
Therefore, there are four cases of operator mixing:\par
(I) Nonresonant diagonalizable $\frac{\gamma_0}{\beta_0}$. \par
(II) Resonant diagonalizable $\frac{\gamma_0}{\beta_0}$. \par
(III) Nonresonant nondiagonalizable $\frac{\gamma_0}{\beta_0}$. \par
(IV) Resonant nondiagonalizable $\frac{\gamma_0}{\beta_0}$. \par
In the case (I), $Z(x,\mu)$ is diagonalizable to all orders of perturbation theory, since the mixing is nonresonant and $\frac{\gamma_0}{\beta_0}$ is diagonalizable.\par
The remaining cases, where $Z(x,\mu)$ is not actually diagonalizable, will be analyzed in a forthcoming paper \cite{BB3}. \par
Specifically, we will work out in \cite{BB3} the canonical form of $ -\frac{\gamma(g)}{\beta(g)}$ for resonant mixing -- that is different from eq. \eqref{0} -- . \par 
In the case (I), the UV asymptotics of $Z(x, \mu)$ reduces essentially to the multiplicatively renormalizable case:
\begin{equation}
\label{01.9}
Z_{i}(x, \mu)=\exp\left(\int^{g(\mu)}_{g(x)}\frac{\gamma_{0i}}{\beta_0 g}dg\right) = \left(\frac{g(\mu)}{g(x)}\right)^{\frac{\gamma_{0i}}{\beta_0}}
\end{equation}
in the diagonal basis, where $Z_i(x,\mu)$ and $\gamma_{0i}$ denote the eigenvalues of the corresponding matrices.\par
Of course, $Z(x, \mu)$ in any other renormalization scheme can be reconstructed from the canonical diagonal form above -- if it exists -- by working out the other way around the appropriate change of basis according to eq. \eqref{Z}.\par
Then, in the case (I) Eq. \eqref{2.2} reads:
\bea \label{10}
G_{ik}(x) &=&  
Z_{i}(x, \mu) \mathcal{G}_{ik}(x,g(\mu),\mu) Z_{k}(x, \mu)
\eea
in the diagonal basis, where no sum on the indices $i,k$ is understood. Eq. \eqref{10} furnishes the UV asymptotics of $G_{ik}(x)$, provided that the asymptotics of $\mathcal{G}_{ik}(x,g(\mu),\mu)$ is known \cite{BB2} as well.\par
We believe that the aforementioned employment of the Poincar\'e-Dulac theorem makes the subject of operator mixing in the physics literature more transparent than in previous treatments \cite{Sonoda}. \par

 \section{Differential geometry of renormalization} \label{3}

We point out that renormalization may be interpreted in a differential-geometric setting, where a (finite) change of renormalization scheme, i.e., a coupling-dependent change of the operator basis:
\bea \label{b}
O'_i(x)=S_{ik}(g) O_k(x)
\eea
is interpreted as a matrix-valued (formal \footnote{A formal series is not assumed to be convergent and, indeed, in the present paper we do not assume that the series in eqs. \eqref{2} and \eqref{1.6} are convergent, since they arise from perturbation theory.}) real-analytic invertible gauge transformation $S(g)$.
Accordingly, the matrix $A(g)$:
\bea \label{A}
A(g)=-\frac{\gamma(g)}{\beta(g)}&=& \frac{1}{g} \left(A_0 + \sum^{\infty}_ {n=1} A_{2n} g^{2n} \right) \nonumber \\
&=& \frac{1}{g} \left(\frac{\gamma_0}{\beta_0} +\cdots \right)
\eea
that occurs in the system of ordinary differential equations defining $Z(x, \mu)$ by eqs. \eqref{1.6} and \eqref{2}:
\bea \label{1.700}
\left(\frac{\partial}{\partial g} +\frac{\gamma(g)}{\beta(g)}\right) Z =0
\eea
is interpreted as a (formal) real-analytic connection, with a simple pole at $g=0$, that for the gauge transformation in eq. \eqref{b} transforms as:
\bea
A'(g)= S(g)A(g)S^{-1}(g)+ \frac{\partial S(g)}{\partial g} S^{-1}(g)
\eea
Morevover,
\bea
\mathcal{D} =  \frac{\partial}{\partial g} - A(g)
\eea
is interpreted as the corresponding covariant derivative that defines the linear system:
\bea \label{ls}
\mathcal{D} X=  \left(\frac{\partial}{\partial g} - A(g)\right) X=0
\eea
whose solution with a suitable initial condition is $Z(x, \mu)$.\par
As a consequence, $Z(x, \mu)$ is 
interpreted as a Wilson line associated to the aforementioned connection:
\bea
Z(x, \mu)=P\exp\left(\int ^{g(\mu)}_{g(x)} A(g) \, dg\right)
\eea
that transforms as:
\bea \label{Z}
Z'(x, \mu)= S(g(\mu)) Z(x, \mu) S^{-1}(g(x))
\eea
for the gauge transformation $S(g)$.\par
Besides, by allowing the coupling to be complex valued, everything that we have mentioned applies in the (formal) holomorphic setting, instead of the real-analytic one.\par
Hence, by summarizing, a change of renormalization scheme is interpreted as a (formal) holomorphic gauge transformation, $-\frac{\gamma(g)}{\beta(g)}$ as a (formal) meromorphic connection with a Fuchsian singularity at $g=0$, and $Z(x,\mu)$ as a Wilson line.

\section{Canonical nonresonant form for $-\frac{\gamma(g)}{\beta(g)}$ by the Poincar\'e-Dulac theorem} \label{4}

According to the interpretation above, the easiest way to compute the UV asymptotics of $Z(x, \mu)$ consists in setting the meromorphic connection in eq. \eqref{A}
in a canonical form by a suitable holomorphic gauge transformation.\par
Specifically, if the nonresonant condition in eq. \eqref{rce}
is satisfied, a (formal) holomorphic gauge transformation exists that sets $A(g)$ in eq. \eqref{A} in the canonical nonresonant form -- the Euler form \cite{PD1} --:
\bea \label{nr}
A'(g)= \frac{\gamma_0}{\beta_0} \frac{1}{g}
\eea
according to the Poincar\'e-Dulac theorem.\par
In this respect, the only minor refinement that we need for applying the Poincar\'e-Dulac theorem to eq. \eqref{A} is the observation that the inductive procedure in its proof \cite{PD1} works as well by only restricting to the even powers of $g$ in eq. \eqref{27} that match the even powers of $g$ in the brackets in the rhs of eq. \eqref{A}.\par
As a consequence, the nonresonant condition in eq. \eqref{rce} only involves positive even integers, as opposed to the general case (section \ref{PD}).

\section{A condensed proof of the Poincar\'e-Dulac theorem for nonresonant diagonalizable $A_0$} \label{PD}

We provide a condensed proof of (the linear version of) the Poincar\'e-Dulac theorem \cite{PD1} for nonresonant diagonalizable $A_0$, which includes the case (I) in the setting of operator mixing for a massless QCD-like theory. \par
The proof in the general case will be worked out in \cite{BB3}.\par
\emph{Poincar\'e-Dulac theorem for nonresonant diagonalizable $A_0$}:\par
The linear system in eq. \eqref{ls}, 
where the meromorphic connection $A(g)$, with a Fuchsian singularity at $g=0$, admits the (formal) expansion:
\bea \label{sys2}
A(g)= \frac{1}{g} \left(A_0 + \sum^{\infty}_ {n=1} A_{n} g^{n} \right)
\eea
with $A_0$ diagonalizable and eigenvalues $\text{diag}(\lambda_1, \lambda_2, \cdots )=\Lambda$, in nonincreasing order $\lambda_1 \geq \lambda_2 \geq \cdots$, satisfying the nonresonant condition:
\bea
\lambda_i -\lambda_j \neq k
\eea
for $i \leq j$ and $k$ a positive integer,
may be set, by a (formal) holomorphic invertible gauge transformation, in the Euler normal form \footnote{In the present paper, we refer to it as the canonical nonresonant diagonal form.}:
\bea \label{canres2}
A'(g)= \frac{1}{g} \Lambda
\eea
We only report the key aspects of the proof, leaving more details to \cite{PD1}. \par
\emph{Proof}:\par
The proof proceeds by induction on $k=1,2, \cdots$ by demonstrating that, once $A_0$ and the first $k-1$ matrix coefficients, $A_1,\cdots,A_{k-1}$, have been set in the Euler normal form above -- i.e., $A_0$ diagonal and $ A_1,\cdots,A_{k-1}=0$ -- a holomorphic gauge transformation exists that leaves them invariant and also sets the $k$-th coefficient, $A_{k}$, to $0$. \par
The $0$ step of the induction consists just in setting $A_0$ in diagonal form -- with the eigenvalues in nonincreasing order as in the statement of the theorem -- by a global (i.e., constant) gauge transformation. \par
At the $k$-th step, we choose the holomorphic gauge transformation in the form: 
\bea \label{27}
S_k(g)=1+ g^k H_k
\eea
with $H_k$ a matrix to be found momentarily. Its inverse is:
\bea
S^{-1}_k(g)= (1+ g^k H_k)^{-1} = 1- g^k H_k + \cdots
\eea
where the dots represent terms of order higher than $g^{k}$.\par
The gauge action of $S_k(g)$ on the connection $A(g)$ furnishes:
\bea \label{ind}
A'(g) &=&  k g^{k-1} H_k ( 1+ g^k H_k)^{-1} +  (1+ g^k H_k) A(g)( 1+ g^k H_k)^{-1} \nonumber \\
&=& k g^{k-1} H_k ( 1+g^k H_k)^{-1}+ (1+ g^k H_k)  \frac{1}{g} \left(A_0 + \sum^{\infty}_ {n=1} A_{n} g^{n} \right) ( 1+ g^k H_k)^{-1} \nonumber \\
&=& k g^{k-1} H_k ( 1- \cdots)+  (1+ g^k H_k)  \frac{1}{g} \left(A_0 + \sum^{\infty}_ {n=1} A_{n} g^{n} \right) (1- g^k H_k+\cdots) \nonumber \\
&=&k g^{k-1} H_k  +    \frac{1}{g} \left(A_0 + \sum^{k}_ {n=1} A_{n} g^{n} \right)+g^{k-1} (H_kA_0-A_0H_k) + \cdots\nonumber \\
&=&  g^{k-1} (k H_k + H_k A_0 - A_0 H_k)+ A_{k-1}(g) + g^{k-1} A_k+ \cdots
\eea
where we have skipped in the dots all the terms that contribute to an order higher than $g^{k-1}$, and we have set:
\bea
A_{k-1}(g) =  \frac{1}{g} \left(A_0 + \sum^{k-1}_ {n=1} A_{n} g^{n} \right)
\eea
that is the part of $A(g)$ that is not affected by the gauge transformation $S_k(g)$, and thus verifies the hypotheses of the induction --  i.e., that $A_1, \cdots, A_{k-1}$ vanish --. \par
Therefore, by eq. \eqref{ind} the $k$-th matrix coefficient, $A_k$, may be eliminated from the expansion of $A'(g)$ to the order of $g^{k-1}$ provided that an $H_k$ exists such that:
\bea
A_k+(k H_k + H_k A_0 - A_0 H_k)= A_k+ (k-ad_{A_0}) H_k=0
\eea
with $ad_{A_0}Y=[A_0,Y]$.
If the inverse of $ad_{A_0}-k$ exists, the unique solution for $H_k$ is:
\bea
H_k=(ad_{A_0}-k)^{-1} A_k
\eea
Hence, to prove the theorem, we should demonstrate that, under the hypotheses of the theorem, $ad_{A_0}-k$ is invertible, i.e., its kernel is trivial. 
\par
Now $ad_{\Lambda}-k$, as a linear operator that acts on matrices, is diagonal, with eigenvalues $\lambda_{i}-\lambda_{j}-k$ and the matrices $E_{ij}$, whose only nonvanishing entries are $(E_{ij})_{ij}$, as eigenvectors. 
The eigenvectors $E_{ij}$, normalized in such a way that $(E_{ij})_{ij}=1$, form an orthonormal basis for the matrices.\par
Thus, $E_{ij}$ belongs to the kernel of $ad_{\Lambda}-k$ if and only if $\lambda_{i}-\lambda_{j}-k=0$. \par
As a consequence, since  $\lambda_{i}-\lambda_{j}-k \neq 0$ for every $i,j$ by the hypotheses of the theorem, the kernel of $ad_{\Lambda}-k$ only contains the $0$ matrix, and the proof is complete.\\

\end{document}